# Vibrational quality classification of metallic turbine blades under measurement uncertainty

Liangliang CHENG [1,2], Vahid YAGHOUBI [1,2], Wim V. PAEPEGEM [1], Mathias KERSEMANS [1]

[1] Mechanics of Materials and Structures (MMS), Ghent University, Technologiepark 46, B-9052,Zwijnaarde, Belgium
Email: liangliang.cheng@ugent.be; vahid.yaghoubi@ugent.be; wim.vanpaepegem@ugent.be; mathias.kersemans@ugent.be
[2] SIM program M3 DETECT-ION, Technologiepark 48 , B-9052 Zwijnaarde, Belgium

**Abstract**
Non-destructive testing on metallic turbine blades is a challenging task due to their complex geometry. Vibrational testing such as Process Compensated Resonance Testing (PCRT) has shown an efficient approach, which first measures the vibrational response of the turbine blades, and then employs a classifier to determine if the quality of the turbine blades are good or bad. Our previous work mainly concentrated on the development of Mahalanobis distance-based classifiers which are fed by the measured vibrational features (such as resonant frequencies). In practice, however, measurement errors could lead to a biased trained classifier, potentially resulting in the wrong quality classifications of the turbine blade. In this study, we investigate the classification problem of turbine blades under measurement uncertainty. For this, the concept of Interval Mahalanobis Space is employed, leading to the Integrated Interval Mahalanobis Classification system (IIMCS) classifier which has high robustness against measurement uncertainty.
The IIMCS employs Binary Particle Swarm Optimization (BPSO) to filter out those resonant frequencies which contribute most to the information in the system. A Monte Carlo Simulation scheme is employed to analyze the sensitivity of the resonant frequencies to the measurement uncertainty. This yields an indicator of reliability, indicating the confidence level of the final classification results. The developed IIMCS methodology is applied to an experimental case study of equiaxed nickel alloy first-stage turbine blades, showing good and robust classification performance.

## 1. Introduction

The Mahalanobis Taguchi System, termed MTS [1], is a pattern recognition method that incorporates the Mahalanobis Distance (MD) [2] designated to differentiate patterns between different groups, which has been applied not only to diagnose the system performance but also to optimize decisions in the system design and implementation.
The implementation of MTS involves two main components: Orthogonal Array (OA) employed to screen the significant variables and Signal-to-Noise Ratio (SNR) used to evaluate the measurement quality. Nevertheless, it serves as a standard diagnostic tool in many industrial fields, still being questioned by many scholars in the feature selection and threshold determination procedures [3-4].
To alleviate the shortcomings of the original MTS approach, the authors thereafter proposed the Mahalanobis Classification System (MCS) [5] and its integrated version (IMCS) [6], subsequently. More specifically, MCS employs Binary Particle Swarm Optimization (BPSO) [7] and PSO/Machine Learning tools (ML) [8-10] to filter out high dependable features and address the threshold determination issues, respectively. Despite the achieved promising outcomes using MCS, it still faces some challenges: a significant number of user-defined parameters and overfitting issues. To handle these challenges, an enhanced version of MCS was designed that incorporates threshold determination as part of the feature selection process. This Integrated Mahalanobis Classification System (IMCS) improved also the classification performance. For more details about MCS and IMCS, readers are referred to [5-6].

However, the presence of measurement uncertainty as a natural attribute in conducting experiments results in biased classifiers in the source data that were not considered in either MCS or IMCS. The present work presents a simple, yet effective, manner to account for measurement uncertainty by using the concept of Interval Mahalanobis Distance (IMD) [11]. This IMD is integrated into the classification scheme, resulting in the Integrated Interval Mahalanobis Classification System (IIMCS). Apart from improving the classification procedure, it also provides an effective indicator for classification reliability.

## 2. Integrated interval Mahalanobis classification system

Despite the demonstrated favorable performance of (I)MCS for classifying metal parts based on their vibration responses, it does not take into account possible measurement uncertainties and provide an assessment of the reliability of the classification results. To achieve this, the (I)MCS classification framework was extended by merging different concepts, resulting in the Integrated Interval Mahalanobis Classification System IIMCS. The general scheme of the IIMCS algorithm is shown in Figure 1.

The main novelties in IIMCS are the implementation of Monte Carlo Method [12] ( termed MCM in Figure 1) to infer the Interval MD and an indicator to point out the classification reliability.

The IMD has been defined as equation (1)

$$\boldsymbol{IMD}_k = [MD_k^1, MD_k^2, \ldots, MD_k^m] \tag{1}$$

In which $m$ and $k$ stand for the MCM total runs and sample index.

In practical applications, the impacts of data uncertainty and learner stability can be dispersed on the final classification results. Benefiting from the concept of IMD, classification reliability can be inferred and considered as an effective tool to indicate the reliability of classification results. The classification reliability is indicated as $r$, ranges from [Good, Bad] = [-1,1]. The plus-minus sign of reliability values indicates the sample classified as Bad (-) and Good (+). And the absolute values of the reliability imply the possibility to classify the sample as Bad or Good. The greater the absolute value of the reliability, the more confident the prediction.

Thus, the reliability of classification is defined as

$$r = \begin{cases} -p^{Good} & if\ p^{Good} > p^{Bad} & r \in [-1,-0.5) \\ p^{Bad} & if\ p^{Good} < p^{Bad} & r \in (0.5, 1] \\ 0 & if\ p^{Good} = p^{Bad} & r = -0.5, 0.5 \end{cases} \tag{2}$$

In which $p^{Good}$ and $p^{Bad}$ are used to represent the assigned probability to the sample being classified as Good and Bad, respectively. The following example in Figure.1 illustrates the concept of classification reliability.

It is assumed that the logarithmic of chi-squared distributed variables is approximately Gaussian distributed and the scaled IMD approximately then corresponds to a Gaussian distribution [5]. Figure 1 indicates that the probability of being classified as Bad and Good is 90% and 10%, respectively, according to equation (2), drawing a conclusion that this part is sorted as Bad with a classification reliability of 0.9 ($r = p^{Bad} = 0.9$).

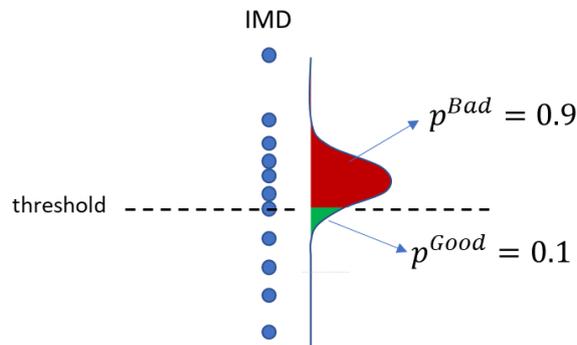

Figure.1 The illustration of classification reliability

Hence, this approach provides an opportunity to investigate how reliable the classification results are.

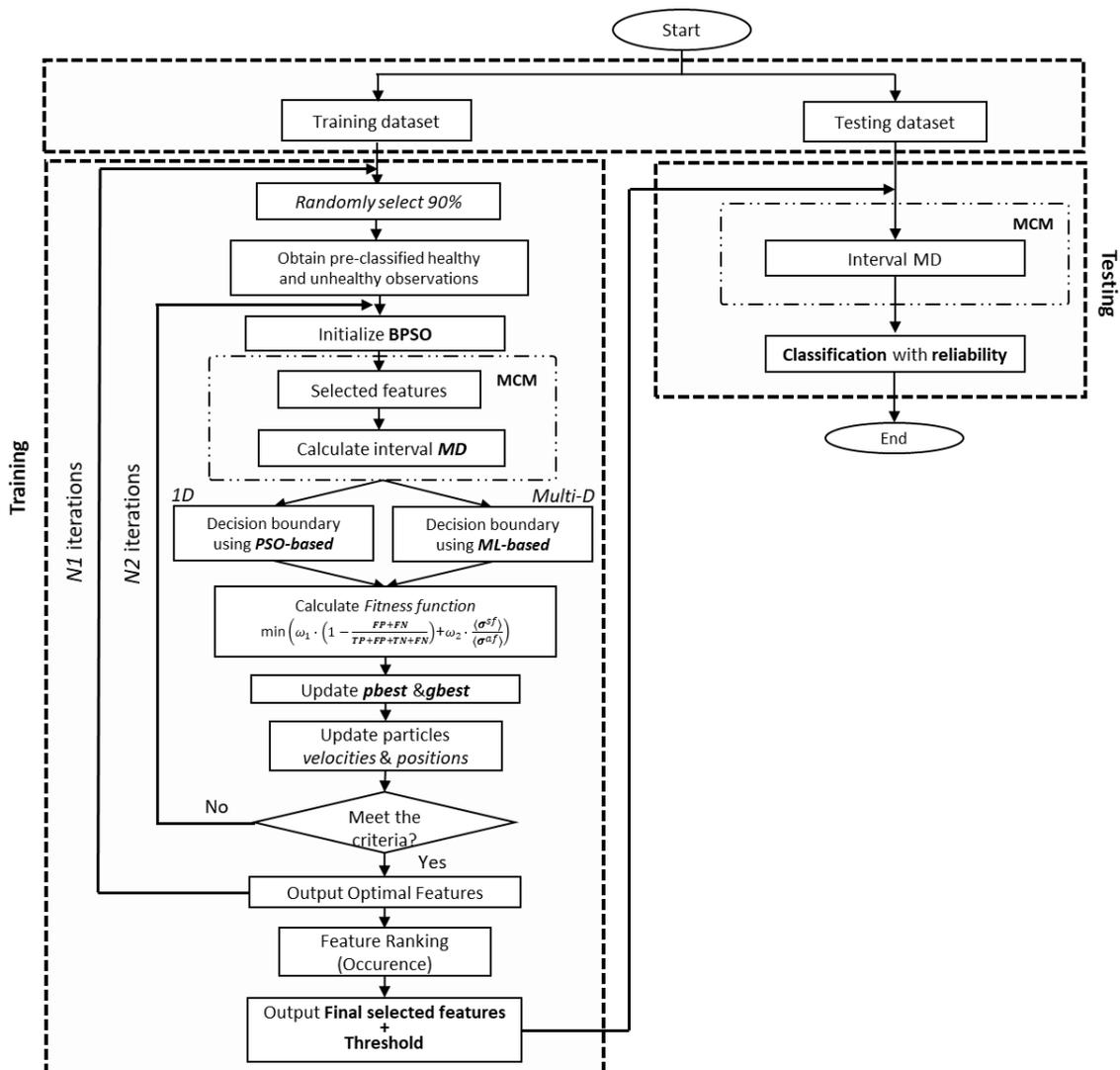

Figure.2 The schematic of IIMCS algorithm

In short, the flowchart of IIMCS in Figure.2 can be briefly described as: The original dataset consists of $n$ samples and $p$ features, which are first randomly divided into training and test datasets. Using the BPSO optimizer with $N2$ iterations, the optimal subset of features is searched in the $m$-run MCM framework to maximize the classification accuracy of the training dataset. Furthermore, the whole operation will be run $N1$ times to further filter out the most reliable features based on their frequency of occurrence. Note that the threshold determination in this study centers only on the one-dimensional MD space, and therefore the PSO optimizer is utilized to exploit the optimal threshold on the basis of the resulting IMD values rather than the single point MD values in both MCS and IMCS.

### 3. A case study on metallic turbine blades

This section shows a case study of an equiaxed nickel alloy primary turbine blade with a generic model and its sectional CAD drawing displayed in Figure 3. A total of 232 blades consisting of 193 healthy and 39 defective blades are available. These defective blades contain several types of defects, such as microstructural changes due to high temperatures; airfoil cracking; intergranular erosion (corrosion); thin walls due to casting; maintenance and repair operations, and service wear.

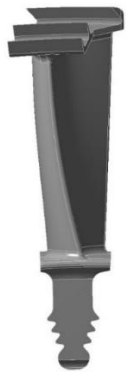
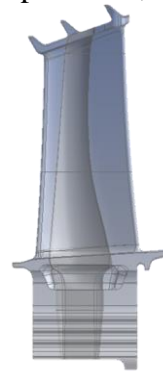

Figure.3a Generic model of the turbine blade     Figure.3b Sectional drawing of the turbine blade

Vibrant Corporation conducted vibration testing of turbine blades in the frequency range of [3, 38] kHz utilizing compensated resonance testing (PCRT) techniques [13]. From the vibration response, 15 resonance frequencies $f_1 \sim f_{15}$ were identified for further analysis. To assess the potential measurement uncertainty or repeatability of the measurements, 31 repeated measurements were performed on the nominal turbine blades.

Figure.4a and Figure.4b present the IMD results based on a complete set of 15 features $f_1 \sim f_{15}$ and the optimized subset of features $(f_2, f_4, f_{14}, f_{15})$, respectively. For better visualization, the Scaled MD [5], which is the log-normalized MD, is used. It is evident from Figure.4 that the margin of IMD values is significantly shrunk via the adoption of the optimal feature subset $(f_2, f_4, f_{14}, f_{15})$.

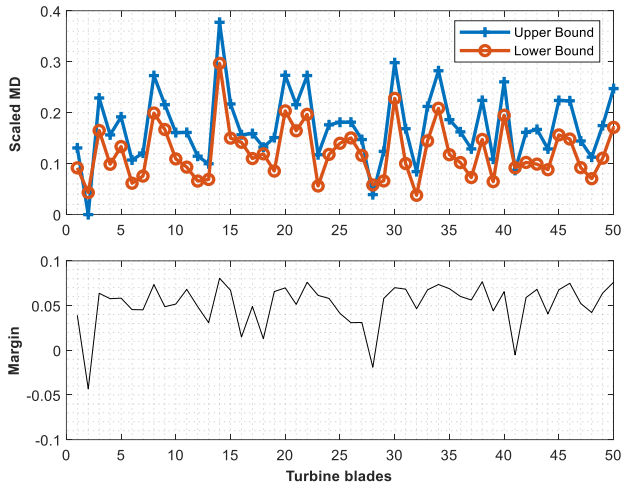
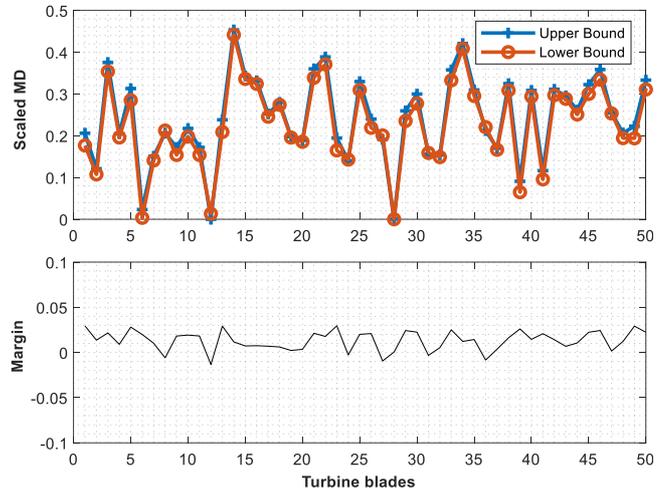

Figure.4a IMD based on the whole fifteen features       Figure.4b IMD based on the selected four features

In addition to having a narrower range of IMD values, the reliability of each test blade shown in Figure 5 also facilitates the determination of the health status of the test turbine blades. Two misclassifications on the testing dataset are observed, resulting in an accuracy of 97.85%. Of interest, the turbine blade 87 is falsely categorized as healthy, but only with a reliability value of $r$ = -0.78.

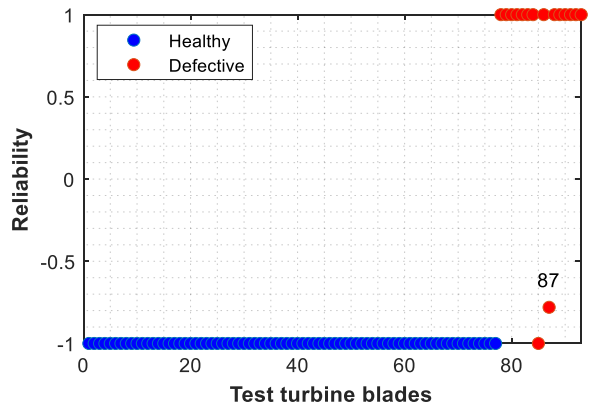

Figure.5 Classification of the testing blades

## 4. Conclusions

An Integrated Interval Mahalanobis Classification System IIMCS is proposed for classifying complex metal parts based on their broadband vibration response. The IIMCS classifier is proposed to address the challenges in dealing with measurement uncertainties in the context of the Mahalanobis-Taguchi System MTS. IIMCS employs a Monte Carlo method framework to propagate the measurement uncertainty into the Mahalanobis distance space MD to obtain the interval MD space and assigns an indicator to each test part intending to evaluate the classification reliability. The proposed IIMCS is demonstrated in an experimental case study of an equiaxed nickel alloy first stage turbine blade, showing satisfactory performance.

**Acknowledgements**
The authors acknowledge the ICON project DETECT-ION (HBC.2017.0603) which fits in the SIM research program MacroModelMat (M3) coordinated by Siemens (Siemens Digital Industries Software, Belgium) and funded by SIM (Strategic Initiative Materials in Flanders) and VLAIO (Flemish government agency Flanders Innovation & Entrepreneurship).
Vibrant Corporation is gratefully acknowledged for providing anonymous datasets of the turbine blades.